\documentclass[%
reprint,
superscriptaddress,
amsmath,amssymb,
aps,
prb,
floatfix,
a4paper
]{revtex4-2}

\usepackage{graphicx}% Include figure files
\usepackage{dcolumn}% Align table columns on decimal point
\usepackage{bm}% bold math
\usepackage{gensymb}
\usepackage{xspace}
\usepackage{textcomp}

\def\cuf{Cu$_2$F$_5$\xspace}

\begin{document}

% \preprint{APS/123-QED}

% \title{Novel cuprate with 2D-exchange interaction: Cu$_2$F$_5$}
% \title{Unusual 2D magnetic interaction between S=1 and S=\textonehalf\xspace ions in Cu$_2$F$_5$}
\title{Mixed spin S=1 and S=\textonehalf\xspace layered lattice in Cu$_2$F$_5$}

\author{Dmitry~M.~Korotin}
	\email{dmitry@korotin.name}
	\affiliation{M.N. Mikheev Institute of Metal Physics of Ural Branch of Russian Academy of Sciences, 18 S. Kovalevskaya St., Yekaterinburg, 620137, Russia.}
	\affiliation{Skolkovo Institute of Science and Technology, 30 Bolshoy Boulevard, bld.1, Moscow, 121205, Russia}

\author{Dmitry~Y.~Novoselov}
	\affiliation{M.N. Mikheev Institute of Metal Physics of Ural Branch of Russian Academy of Sciences, 18 S. Kovalevskaya St., Yekaterinburg, 620137, Russia.}
	\affiliation{Skolkovo Institute of Science and Technology, 30 Bolshoy Boulevard, bld.1, Moscow, 121205, Russia}
    \affiliation{Department of Theoretical Physics and Applied Mathematics, Ural Federal University, 19 Mira St., Yekaterinburg 620002, Russia}

\author{Vladimir~I.~Anisimov}
	\affiliation{M.N. Mikheev Institute of Metal Physics of Ural Branch of Russian Academy of Sciences, 18 S. Kovalevskaya St., Yekaterinburg, 620137, Russia.}
	\affiliation{Skolkovo Institute of Science and Technology, 30 Bolshoy Boulevard, bld.1, Moscow, 121205, Russia}
    \affiliation{Department of Theoretical Physics and Applied Mathematics, Ural Federal University, 19 Mira St., Yekaterinburg 620002, Russia}

\author{Artem~R.~Oganov}
\affiliation{Skolkovo Institute of Science and Technology, 30 Bolshoy Boulevard, bld.1, Moscow, 121205, Russia}

\date{\today}

\begin{abstract}
The electronic and magnetic structure, including the Heisenberg model exchange interaction parameters, was explored for the recently proposed novel cuprate \cuf. Using the DFT+U calculation, it is shown that the compound is formed by two types of copper ions with $d^9$ and $d^8$ electronic configurations. 
We have found a very stable antiferromagnetic ordering with strong anisotropy of exchange interaction that results in the appearance of an unusual 2D-magnetism: within the (100)-plane the exchange between the S=1 and S=1/2 Cu ions has almost the same strength as between the two S=1 ions. The interplane magnetic interaction is five times weaker than the in-plane one.
\end{abstract}

%\keywords{Suggested keywords}%Use showkeys class option if keyword
                              %display desired
\maketitle

\section{Introduction}

The low-dimensional spin-lattices exist in a plethora of forms such as spin-ladders~\cite{Gopalan1994,Notbohm2007,PhysRevLett.83.1387}, plaquettes~\cite{Zayed2017}, dimers and zig-zags~\cite{Ueda1998,Mostovoy1999}. The compounds owning these magnetic structures demonstrate unusual magnetic excitation spectra, including the appearance of the spin-gap.  In the present work, we've found a new type of 2D spin-lattice in \cuf -- the novel stable copper fluoride \cuf predicted theoretically recently in our work~\cite{Rybin2021}.
Here we analyze its electronic and magnetic structure in detail, relying on the known crystal structure.

% The \cuf have common structural and electronic structure motifs with high-$T_c$ cuprates, such as La$_2$CuO$_4$ and with prototypical correlated perovskite compound KCuF$_3$: there are Cu$^{2+}$ ions in the center of fluorine octahedra which share their corners. Following the analogy to the mentioned compounds, one can assume the existence in \cuf of AFM ordered spin-lattice of the Cu ions moments. At the same time type of the ordering and dimension of the spin-lattice is unclear and should be obtained in calculations until experimental data appears.

The novel copper fluorides \cuf and CuF$_3$, proposed earlier~\cite{Rybin2021} have structural and electronic similarity with high-$T_c$ cuprates, such as La$_2$CuO$_4$ and with prototypical correlated perovskite compound KCuF$_3$. Following the analogy to  the mentioned  compounds, one can assume  the existence in \cuf of AFM ordered spin-lattice  of the  Cu ions moments. At  the  same time, the type of  the ordering  and dimensionality of  the  spin-lattice  is unclear and  should  be  obtained in  calculations until experimental data appear.

The proposed analogy between \cuf and La$_2$CuO$_4$ inspires us to analyze the spin state of Cu ions in \cuf and to determine the orbital content of the electronic states right below the Fermi level. We have also explored the magnetic exchange interactions within the \cuf crystal to find some 2D magnetic structures. 

\begin{figure}
	\includegraphics[width=0.9\columnwidth]{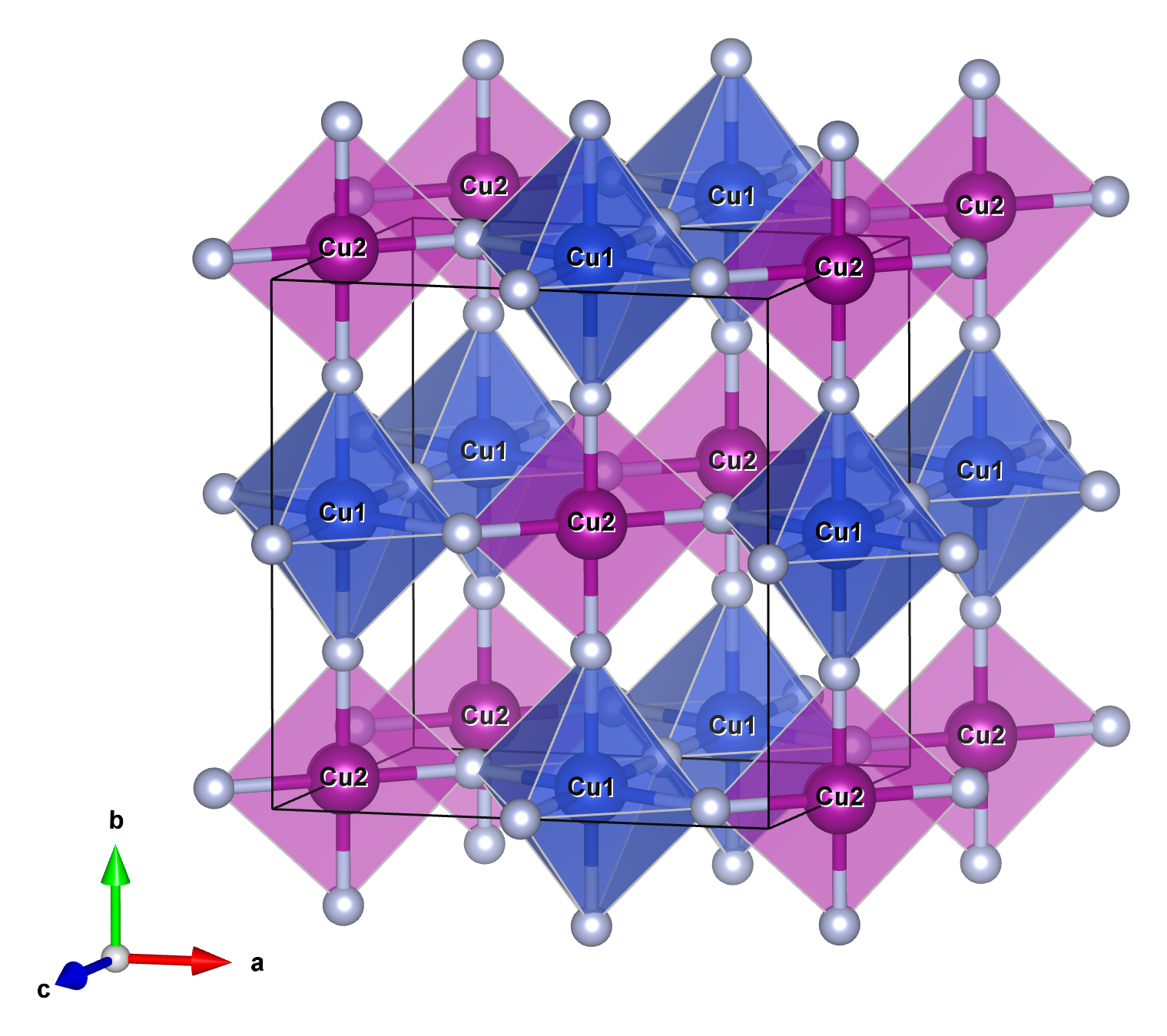}
	\caption{Crystal structure of \cuf. Blue spheres denote Cu ions inside the ligand's octahedron, magenta spheres denote Cu ions in the center of plaquettes, grey spheres -- F ions.}
	\label{fig:struc}
\end{figure}

The  theoretically predicted ~\cite{Rybin2021} (using DFT)  crystal structure of \cuf has monoclinic C2/m symmetry and contains Cu ions with two types of coordination (see Fig.~\ref{fig:struc}). The first type of Cu ions (Cu1) is in the center of a slightly distorted octahedron with three pairwise Cu-F distances (1.98, 1.96, 1.89~$\AA$) and only one F-Cu-F angle equal to 91.3$\degree$ (other F-Cu-F angles are 90$\degree$). Another type of Cu ions (Cu2) is in the center of a square formed by four F ions with two different Cu-F distances (1.83 and 1.86~$\AA$). All F-Cu-F angles within such square plaquette are equal to 90$\degree$. Cu1 and Cu2 ions alternate along $a$ and $b$ crystal axes and form chains of the same type ions along the $c$ axis.

If one neglects (not in calculations, but for simplicity of interpretation) the deviation of the F-Cu1-F angle from the 90$\degree$ value then the Cu1 ion type has the D$_{4h}$ point group symmetry. It means that the Cu $d$-level splits into the $t_{2g}$ ($d_{zx}$, $d_{zy}$, $d_{xy}$) and $e_g$ ($d_{x^2-y^2}$, $d_{3z^2-r^2}$) subshells. 
For the Cu2 type of ions the square-planar surrounding splits the $d$-level into (from highest to lowest): $b_{1g} (d_{x^2-y^2})$, $b_{2g} (d_{xy})$, $e_g (d_{zx}, d_{zy})$ and $a_{1g} (d_{3z^2-r^2})$ subshells.

To interpret our results in terms of Cu $d$-orbitals, we have defined the local coordinate system (LCS) for each Cu ion as shown in Fig.~\ref{fig:exchanges_path} with the $z$-direction perpendicular to the plaquette plane for the Cu2 ion and the $z$-direction along with the crystal $b$-vector for the Cu1 ion. Below, analyzing densities of states, hopping integrals, exchange interaction, {\em etc.} we will refer to the $d$-orbitals defined in this LCS.

\section{Methods}\label{sec:methods}

All calculations were performed using Quantum-ESPRESSO~\cite{Giannozzi2009} package with pseudopotentials from pslibrary set~\cite{pslibrary}. The exchange-correlation functional was chosen to be in Perdew-Burke-Ernzerhof~\cite{PBE} form. The energy cutoff for plane wave wavefunctions and charge-density expansion has been set to 50~Ry and 400~Ry, respectively. Integration in the reciprocal space was done on a regular $8\times8\times8$ $k$-points mesh in the irreducible part of the Brillouin zone. 

In  Section~\ref{sec:results} we first explore the electronic structure  within  DFT and then,  keeping  in mind the  partially  filled $d$-shell  of the Cu  ions, continue with the  DFT+U approximation.  The DFT+U method  is a  compromise  between the ability to  describe correlated  $d$-states  and computational cost of the calculations. At each value of Hubbard’s $U$ parameter we relaxed the crystal structure. At the same time, Hund’s parameter $J$ was kept fixed at 0.9~eV in all DFT+U calculations; the relation between  crystal field splitting  of the $e_g$ level of Cu1 ion and $J$ parameter could, in principle, affect  the stability of  the  $S = 1$ magnetic configuration, but we do not expect such  a case  because the splitting of $e_g$ levels is small. $J$ = 0.9~eV is typical  for  cuprates~\cite{Blaha2005,Leonov2008}. As soon as experimental crystal structure of \cuf  will become available, it will be intriguing to  calculate the $U$ and $J$ parameters  with  the  constrained DFT~\cite{hamilt}  or  linear response~\cite{Cococcioni2005}  methods  and investigate the electronic  structure of this compound with a more  advanced  approach such as DFT+DMFT~\cite{DFT+DMFT,Held2006}.

% \textbf{In Section~\ref{sec:results} we first explore electronic structure within DFT and then, keeping in mind the partially filled $d$-shell of the Cu ions, continue with the  DFT+U approximation.

% Since there is no experimental crystal structure of \cuf yet, we were obliged to perform cell relaxation for several Hubbard $U$ parameter values, varying the strength of electronic correlations. The DFT+U method is a compromise between the ability to describe correlated $d$-states and computational heaviness in such relaxations.

% The effect of Hund's $J$ parameter alteration was not explored in the present work. The relation between crystal field splitting of the $e_g$ level of Cu1 ion and $J$ parameter could, in principle, affect the stability of the $S=1$ magnetic configuration. We don't expect such a case because the $e_g$ levels splitting is small. That is why we kept the $J$ parameter fixed equal to 0.9~eV in all DFT+U calculations. This value is typical for cuprates~\cite{Blaha2005,Leonov2008}.

% As soon as an experimental crystal structure of \cuf is available, it is intriguing to calculate the $U$ and $J$ parameters with the constrained DFT~\cite{hamilt} or linear response~\cite{Cococcioni2005} method and investigate the electronic structure of the compound with a more advanced approach such as DFT+DMFT~\cite{DFT+DMFT,Held2006}.}

The convergence criteria used for the crystal cell relaxation within DFT+U are: total energy $< 10^{-6}$~Ry, total force  $< 10^{-3}$~Ry/Bohr, pressure $< 0.5$~kbar.

\section{\label{sec:results} Results }

\begin{figure}
	\includegraphics[width=0.8\columnwidth]{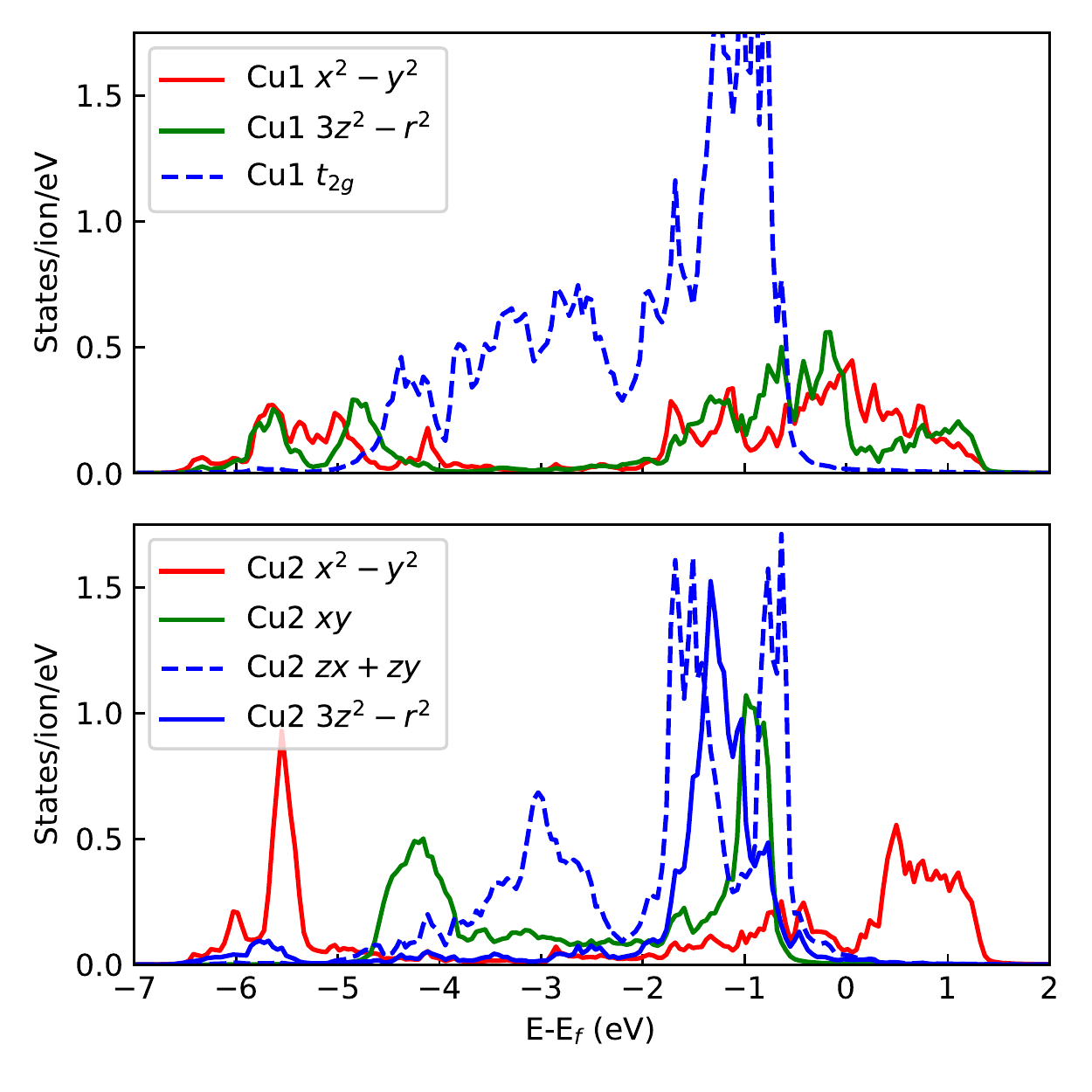}
	\caption{Partial densities of states for \cuf obtained within spin-unpolarized DFT calculation.}
	\label{fig:dft_doses}
\end{figure}

Partial densities of states (pDOSes)  obtained within spin-unpolarized DFT calculation are shown in Fig.~\ref{fig:dft_doses}. 
For the Cu1 ions, there is the filled $t_{2g}$ energy band with the width $\approx$4.5~eV. The same three orbitals ($d_{xy}$, $d_{zx}$ and $d_{zy}$,) are filled for the Cu2 ions too, and the corresponding energy bands are located in the same energy region.

The situation is different for the other two $d-$states. 
Both $e_g$-states of the Cu1 ions are partially filled, their energy bands cross the Fermi level, and the occupations (from DFT) are 0.8~$e$ and 0.74~$e$. Contrary, the $d_{3z^2-r^2}$ orbital of the Cu2 ion (in the center of the flat plaquette) is almost filled, and the corresponding peak of DOS is located $\approx$1.2~eV below the Fermi level. And the $d_{x^2-y^2}$-orbital of the Cu2 ions is partially filled with the occupation 0.62~$e$.

\begin{figure}
	\includegraphics[width=\columnwidth]{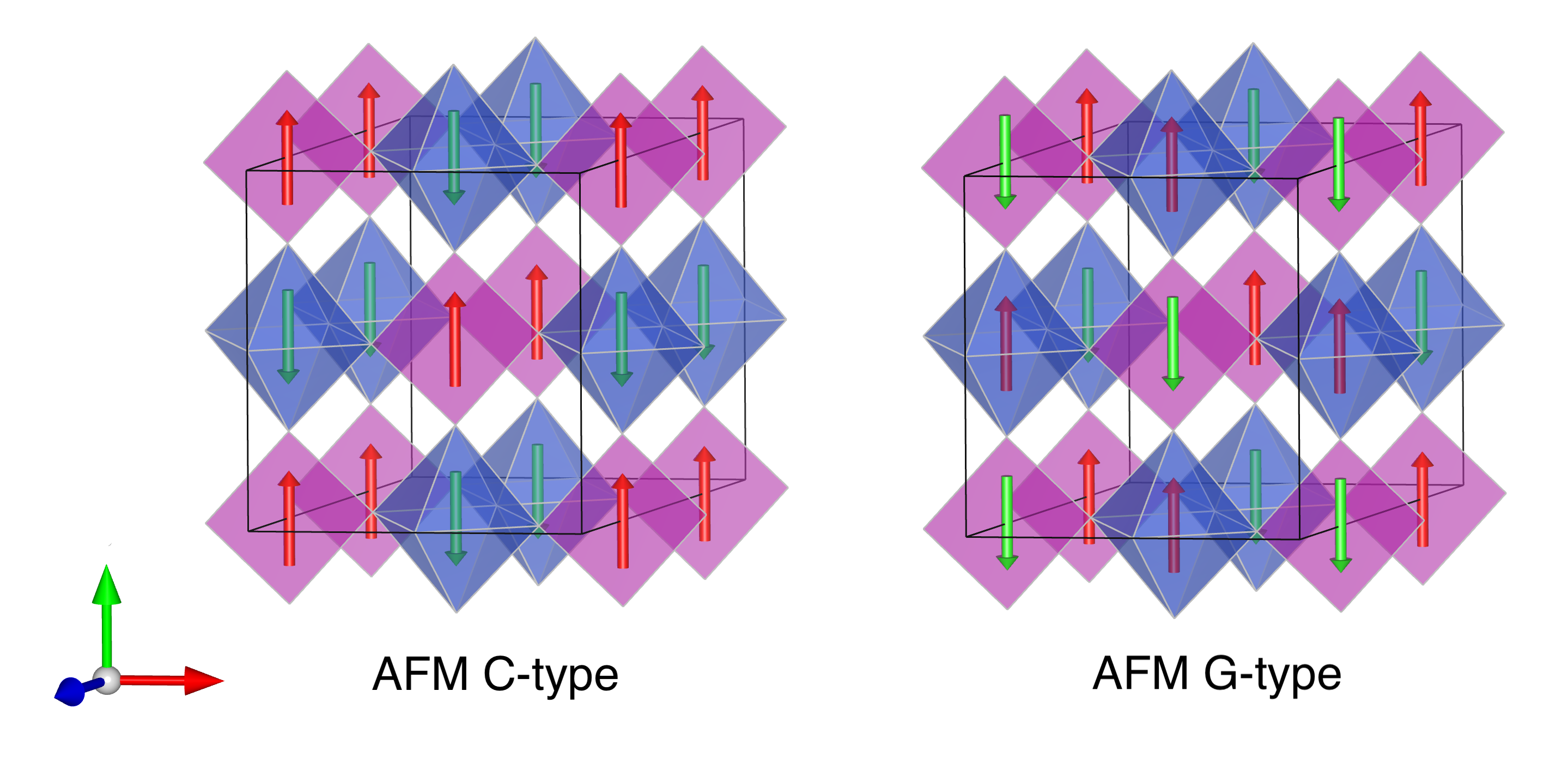}
	\caption{Two types of considered AFM orderings of Cu ions magnetic moments in Cu$_2$F$_5$. Dark blue octahedra are CuF$_6$ octahedra, light-magenta rhombuses are CuF$_4$ plaquettes.}
	\label{fig:afm_orderings}
\end{figure}

Keeping in mind the existence of partially filled $d$-states, one should consider possible magnetic configurations. Also, the strong electronic correlations between the Cu electrons better be taken into account. We have compared the total energies of the Cu$_2$F$_5$ cell calculated for ferromagnetic (FM) and two types of antiferromagnetic (AFM) orderings (shown in Fig.~\ref{fig:afm_orderings}). Since this compound has not yet been made in an experiment, and it is not known is it metallic or insulating, we used various values for the Hubbard $U$ parameter from 2 to 8~eV. The AFM G-type of ordering is always favorable and has the total energy $\approx$300~meV lower than the other two phases. 

\begin{figure*}
	\includegraphics[width=0.8\textwidth]{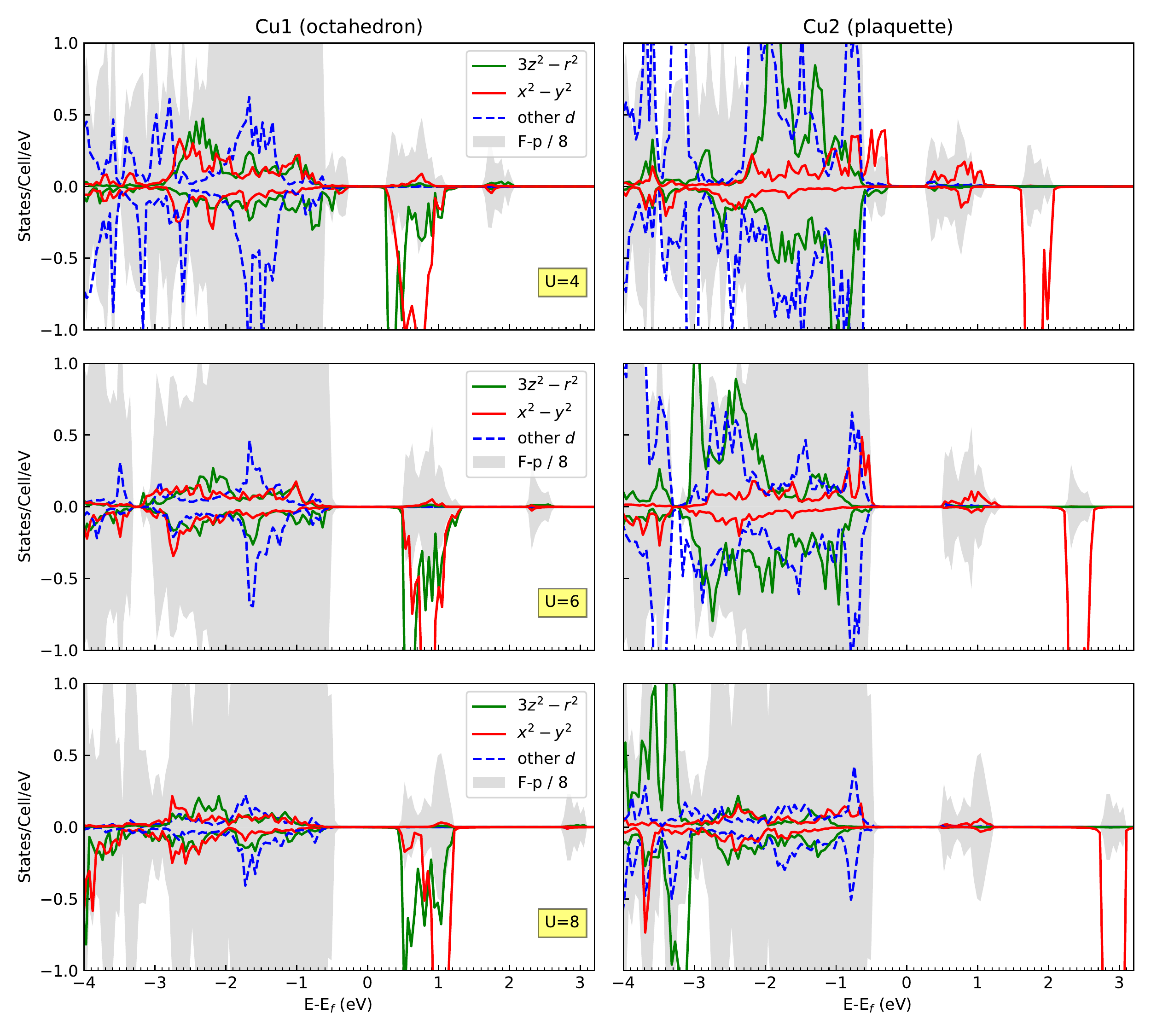}
	\caption{Partial densities of states for \cuf obtained within DFT+U for the AFM-G ordered phase. Positive/negative pDOSes correspond to spin-up/down states respectively.}
	\label{fig:dftu_doses}
\end{figure*}

The partial densities of states for AFM-G ordered magnetic phase of the Cu$_2$F$_5$ calculated within the DFT+U approach are shown in Fig.~\ref{fig:dftu_doses}. For each value of the $U$ parameter, the cell volume and atomic positions were relaxed to obtain the ground-state crystal structure. Starting from $U$=4~eV the \cuf is an insulator. The band gap at $U$=4~eV is 0.48~eV and it broadens with increasing of the $U$ parameter. For this $U$ value it is an antiferromagnetic insulator with the top of the valence band formed by $x^2-y^2$ states of the Cu2 ion. The increase of  $U$ results in the shifting of the occupied Cu-$d$ states into the fluorine band. For $U$=6 and 8~eV,  \cuf is a charge-transfer insulator with the band gap equals to 0.9~eV.

From the left panel of Fig.~\ref{fig:dftu_doses}, it is seen that the Cu ion inside the fluorine octahedron (Cu1 ion) has unusual Cu$^{3+}$ valence with electronic configuration $d^8$. The two peaks in spin-down DOSes for $3z^2-r^2$ and $x^2-y^2$ orbitals at +0.9~eV should be interpreted as an empty $e_g$-states of Cu and the $t_{2g}$ states are filled. The second type of Cu ion (in the center of fluorine plaquettes) has the $d^9$ electronic configuration with an empty spin-down $x^2-y^2$ orbital located at 2-3~eV above the top of the valence band.
Consequently, there is S=1 spin-state for Cu1 ion and S=1/2 spin-state for the plaquette-centered Cu ion.
The position of the bottom of the conduction band is only slightly affected by the electron-electron interaction power due to strong hybridization between the F-$p$ and Cu-$d$ states in the corresponding energy interval.

Using Green's function method based on magnetic-force linear response theory~\cite{exchanges} we computed the Heisenberg exchange interaction between Cu ions up to the 9th nearest neighbor. 
The model Hamiltonian has the form:
\begin{equation}
\label{LP}
H = -\sum_{\langle ij\rangle} J_{ij} {\bf e}_i {\bf e}_j,
\end{equation}
where 
${\bf e}_i$ are the unit vectors pointing in the direction of the $i$th site magnetization, and the summation runs once over each ion pair. 

\begin{figure}
	\includegraphics[width=\columnwidth]{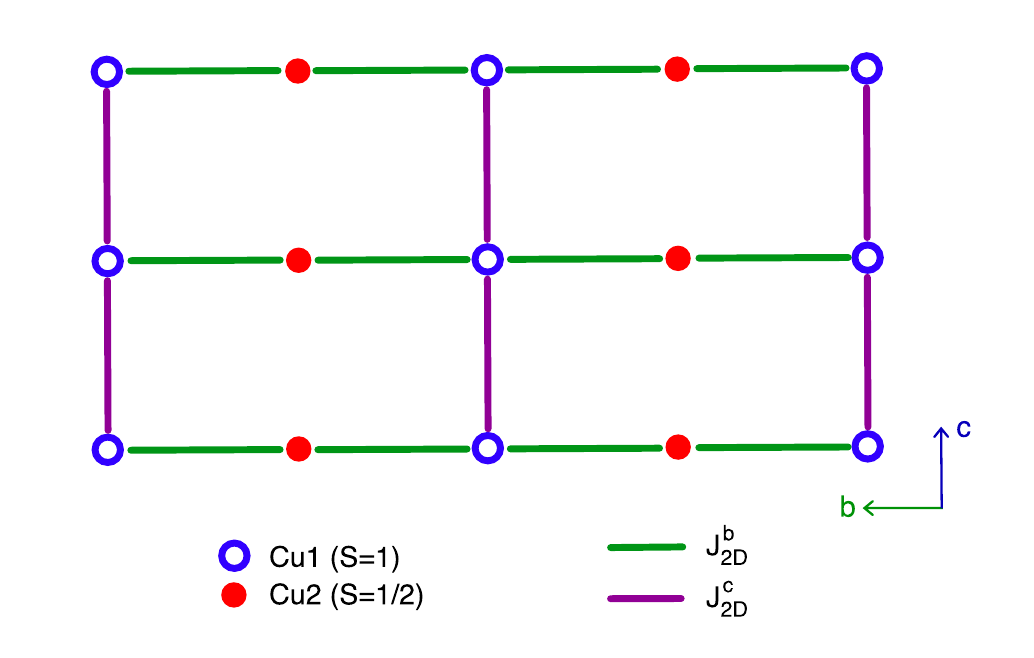}
	\caption{The pattern of the exchange interaction within the (100) layer of Cu ions. Open circles denote Cu1 ions with S=1 in the center of CuF$_6$ octahedron; filled circles are Cu2 ions with S=1/2 (plaquettes centered). The strongest exchanges are: $J_{2D}^b$ (green lines) and $J_{2D}^c$ (violet lines). The interlayer exchange is $J_{\bot}$ is five times smaller than intralayer ones. Other exchange interactions are negligible. Fluorine ions are not shown for clarity.}
	\label{fig:exchanges}
\end{figure}

The obtained $J_i$ values are presented in Table~\ref{tab:exchanges} (only values larger than 1.2~meV are included). The spatial illustration of the exchange interaction directions is shown in Fig.~\ref{fig:exchanges}.

\begin{table}[h!]
\centering
 \begin{tabular}{c  c  c  c} 
 \hline\hline
 $U$ & $J_{2D}^b$~(meV) & $J_{2D}^c$~(meV) & $J_{\bot}$~(meV) \\ [0.5ex] 
 \hline
 4 eV & -32.5 & -34.2 & -6.8 \\ 
 6 eV & -34.5 & -40.2 & -6.9 \\
 8 eV & -32.7 & -42.8 & -6.3 \\
 \hline\hline
\end{tabular}
\caption{Calculated values of exchange interaction parameters within the (100) layer ($J_{2D}^b$ and $J_{2D}^c$) and between these layers ($J_{\bot}$)}
\label{tab:exchanges}
\end{table}

\begin{figure}
	\includegraphics[width=\columnwidth]{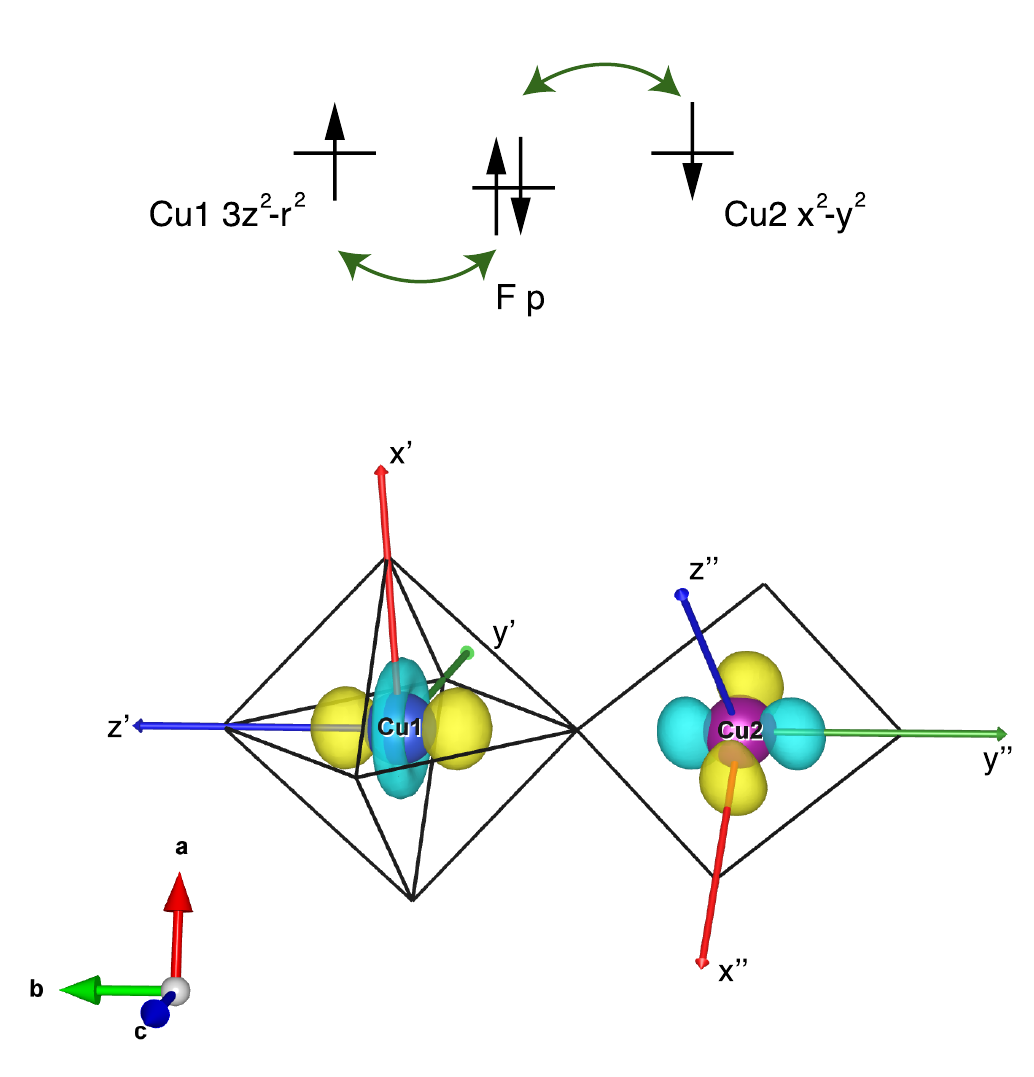}
	\caption{Upper panel: The scheme of the half-filled Cu $d_{3z^2-r^2}$ → F-p → the half-filled Cu $d_{x^2-y^2}$ superexchange mechanism. Lower panel: Two half-filled $d$-orbitals providing the AFM  superexchange interaction along the $b$ crystal axis ($J_{2D}^b$). The $d$-orbitals are defined in own local coordinate systems for the octahedron (x'y'z') and plaquette (x''y''z''). The fluoride $p$-orbital, that mediates the superexchange interaction, located between the $d$-orbitals isn't drawn for picture simplicity.}
	\label{fig:exchanges_path}
\end{figure}

All $J$ values have the same sign that corresponds to the AFM exchange. Both: the Cu1-Cu2 interaction along the $b$-axis ($J_{2D}^b$) together with Cu1-Cu1 interaction along the $c$-axis ($J_{2D}^c$), are five times larger than the exchange interaction along the $a$-crystal axis ($J_{\bot}$). 
The superexchange interaction along the $b$ crystal axis ($J_{2D}^b$) is provided by the hopping of electrons between the half-filled Cu1 $3z^2-r^2$ orbital, the $p$-orbital of fluorine (located between the Cu ions), and the half-filled Cu2 $x^2-y^2$ orbital as illustrated in Fig.~\ref{fig:exchanges_path}. The second half-filled $e_g$ orbital of the Cu1 ion -- $x^2-y^2$ maintains Cu1-F-Cu1 superexchange interaction along the $c$ crystal axis ($J_{2D}^c$). In both interactions described above the Cu-F-Cu angle is 180$^o$. Since this bond angle defines the $p$-$d$ orbitals overlap it, consequently, determines the exchange strength. The Cu1-F-Cu2 along $a$ angle is 129$^o$ only. That sets the much weaker exchange interaction in the direction of the crystal $a$-axis ($J_{\bot}$). As for the Cu2-Cu2 exchange along the $c$-axis, it is negligible because no fluorine ions are providing the superexchange path. 

Thereby a strong superexchange anisotropy in \cuf exists in a way when the in-plane magnetic interactions are significantly larger than the interplane one. 
The absolute values of the exchange parameters are comparable with the superexchange in 2D~\cite{Kastner1998,Mizuno1998}, ladder cuprates~\cite{Mizuno1998} and  1D chain cuprates~\cite{Mizuno1998,Ami1995}. The unusual thing is that the exchange energy between Cu ions with S=1 is almost the same as between the S=1 and S=\textonehalf ions.

%If one considers CuF$_2$~\cite{Miller2020} -- fluorite with Cu ions in the same $d^9$ electronic configuration, but without the CuF$_6$ octahedra in the structure (only CuF$_4$ square plaquettes are there), then the AFM exchange interaction between the nearest Cu ions is an order smaller in CuF$_2$ than in \cuf.

%comparable with the superexchange in silver difluoride AgF$_2$~\cite{Miller2020} ($J^{Cu_2F_5}_{2D} \approx \frac{2}{3} J^{AgF_2}_{2D}$) and the same order of magnitude as in La$_2$CuO$_4$~\cite{Kastner1998}. 

One can suggest that doping of \cuf with electrons would result in filling the Cu1 $3z^2-r^2$ orbital since the bottom of the conduction band is formed by these states according to Fig.~\ref{fig:dftu_doses}. That will shift the Cu1 ion from the $d^8$ configuration and from the half-filled $e_g$-subshell. According to the Goodenough-Kanamori rule~\cite{Goodenough2008}, it will suppress the $J_{2D}^b$ AFM superexchange mechanism illustrated in the upper panel of Fig.~\ref{fig:exchanges_path}. As a result, 1D magnetic chains could arise along the $c$-crystal axis.

% \section{Conclusion}

In conclusion, we have presented a DFT+U study of the electronic and magnetic structure of the novel copper fluoride. The Cu ions in this compound have two different valences: 2+ for Cu ion in the square coordination and 3+ for Cu in octahedral coordination. We have tested various Hubbard interaction parameters $U$ for \cuf and showed that the compound becomes an insulator starting from $U$=4~eV. The value of the energy gap depends only slightly on the $U$ value.

Calculated values of superexchange interaction parameters indicate that the significant magnetic interaction anisotropy exists in \cuf. The new 2D spin-lattice is obtained: the exchange interactions between Cu ions in the (100) planes are five times larger than along the $a$-crystal axis, and within the layer, the exchange between ions with different spins (S=1 and S=\textonehalf) has the same magnitude.

\section*{Acknowledgments}
The presented results were obtained with support of Russian Science Foundation (Project 19-12-00012).

\bibliography{main}

\end{document}